# High-Pressure Tuning of Electrical Transport in Freestanding Oxide Films

Jingxin Chen[1,2,#], Xiang Huang[1,2,#], Zhihan Qiao[1,2], Jiao Li[1,2], , Jiahao Xu[1,2], Haiyang Zhang[1,2], Deyang Li[1,2], Enyang Men[1,2], Hangtian Wang[1,2], Han Zhang[1,3], Jianyu Xie[4], Guolin Zheng[1], Mingliang Tian[1,*], Qun Niu[1,*], Lin Hao[1,*]

[1]Anhui Provincial Key Laboratory of Low-Energy Quantum Materials and Devices, High Magnetic Field Laboratory, HFIPS, Chinese Academy of Sciences, Hefei, Anhui 230031, China

[2]Science Island Branch of Graduate School, University of Science and Technology of China, Hefei 230026, China

[3]School of Microelectronics and Control Engineering, Changzhou University; Changzhou, Jiangsu 213001, China

[4]Spin-X Institute, School of Chemistry and Chemical Engineering, South China University of Technology, Guangdong 511442, China

---

[#]These authors contribute equally. *To whom all correspondence should be addressed. tianml@hmfl.ac.cn, qniu@hmfl.ac.cn, haolin@hmfl.ac.cn.

**Abstract**

Electrical transport in oxide thin films under high pressure remains largely unexplored due to the lack of a universal experimental strategy. Here we develop an approach that enables high-pressure transport measurements in freestanding oxide films by enhancing their mechanical robustness and integrating them with nanoscale high-pressure devices. As a demonstration, we investigate the resistivity of perovskite $SrIrO_3$ films under hydrostatic pressure and uncover a pressure-driven semimetal-insulator transition near 2.5 GPa, followed by an insulator-metal transition around 9 GPa. In the monolayer limit, $SrIrO_3$ remains insulating and robust against pressure up to 5.5 GPa. The contrasting pressure-dependent phase diagrams of three- and two-dimensional iridates reveal a strong interplay between dimensionality and hydrostatic pressure in correlated oxides. Our work establishes a general platform for exploring pressure-driven phenomena in low-dimensional quantum materials.

## Introduction

Oxide thin films represent a large class of quantum materials exhibiting enormously intriguing physical properties, such as superconductivity, versatile long-range (anti-)ferroic orders, and quantized states[1-4]. The strong chemical bonding with the underlying substrate renders oxide thin films elegant platforms for designing and tailoring metastable phases that cannot be obtained in bulk materials[5,6]. The millimeter-scale substrate, on the other hand, significantly shapes the way external stimuli can be applied. For example, it is much easier to apply an in-plane electrical current to oxide thin films than to apply an out-of-plane current, because of the insulating nature of most commercially available substrates. In comparison, there is no restriction on the current direction in a bulk material, as schematically shown in Fig. 1(a). In contrast to the partial restriction on the application of electrical current, there is essentially no effective route to impose high hydrostatic pressure on oxide thin films.

This technical challenge originates from the dilemma in the field of thin films, where a substrate is indispensable for preparing high-quality thin films, but the millimeter-thick sample cannot be fitted into any portable high-pressure apparatus capable of generating hydrostatic pressures higher than 3 GPa[7]. In other words, to accommodate the sample size of thin films, the pressure apparatus must be at least one order of magnitude larger (Fig. 1(b)), which is unrealistic, especially for delicate setups such as the diamond anvil cell (DAC)[8]. Moreover, because of the millimeter-thick substrate, the in-plane force acting on thin films is only $10^{-3}$~$10^{-4}$ of the total force (Fig. 1(b)). It is even more difficult to investigate electrical transport properties due to additional technical challenges in establishing electrical connections under high pressure. As a result, in contrast to the numerous high-pressure studies on bulk materials[9,10], high-pressure (above 3 GPa) investigations of electrical properties in thin films are generally unfeasible, except for a few studies employing large apparatuses[11].

In recent decades, numerous studies have focused on detaching oxide thin films from their underlying substrates by employing soluble buffer layers during film growth[12]. The resulting freestanding films have been shown to exhibit comparable or even enhanced functionalities (*e.g.*, crystalline elasticity) relative to their epitaxial counterparts[13]. Standard lift-off techniques have been developed to obtain freestanding oxide films down to the ultrathin limit of a single unit cell[14]. More recently, versatile sacrificial materials have enabled the fabrication of large-area freestanding films for a wide range of non-ferroelectric oxides[15,16]. The spatial decoupling from the substrate also opens new avenues for externally manipulating oxide thin films[17]. For example, Hong *et al.* achieved continuously tunable strain by uniaxially or biaxially stretching freestanding $La_{0.7}Ca_{0.3}MnO_3$ films, reaching extreme tensile strain[18].

In this work, we develop a universal strategy to apply high hydrostatic pressure to freestanding oxide thin films. To suppress cracking during high-pressure measurements, the films are encapsulated between nanometer-thick ferroelectric (FE) layers. Using a specially designed diamond anvil cell (DAC), hydrostatic pressures up to 16.5 GPa are achieved on freestanding films. With this approach, we uncover pressure-driven semimetal-insulator and insulator-metal transitions in bulk-like $SrIrO_3$ (SIO) films. By atomically controlling the stacking sequence, we further demonstrate that the dimensionality-driven insulating state in monolayer SIO remains robust under high pressure. Beyond enabling studies down to the single-unit-cell limit, we also show that this strategy is broadly applicable to a wide range of oxide thin films.

## Results and discussion

### Encapsuled freestanding films and high-pressure cell

Among the numerous technical challenges in handling freestanding oxide films, the most critical one is the aggravated cracking after multi-step releasing and transferring procedures[19,20]. More importantly, the fragile nature of freestanding oxide films makes it difficult to preserve a complete sample under high pressure, which is crucial for electrical transport measurements. To reduce cracking and improve the universality of the high-pressure strategy, we designed a three-layer structure in which the oxide thin film is encapsulated by two FE blocks, taking advantage of the excellent elasticity and good insulating properties of FE compounds.

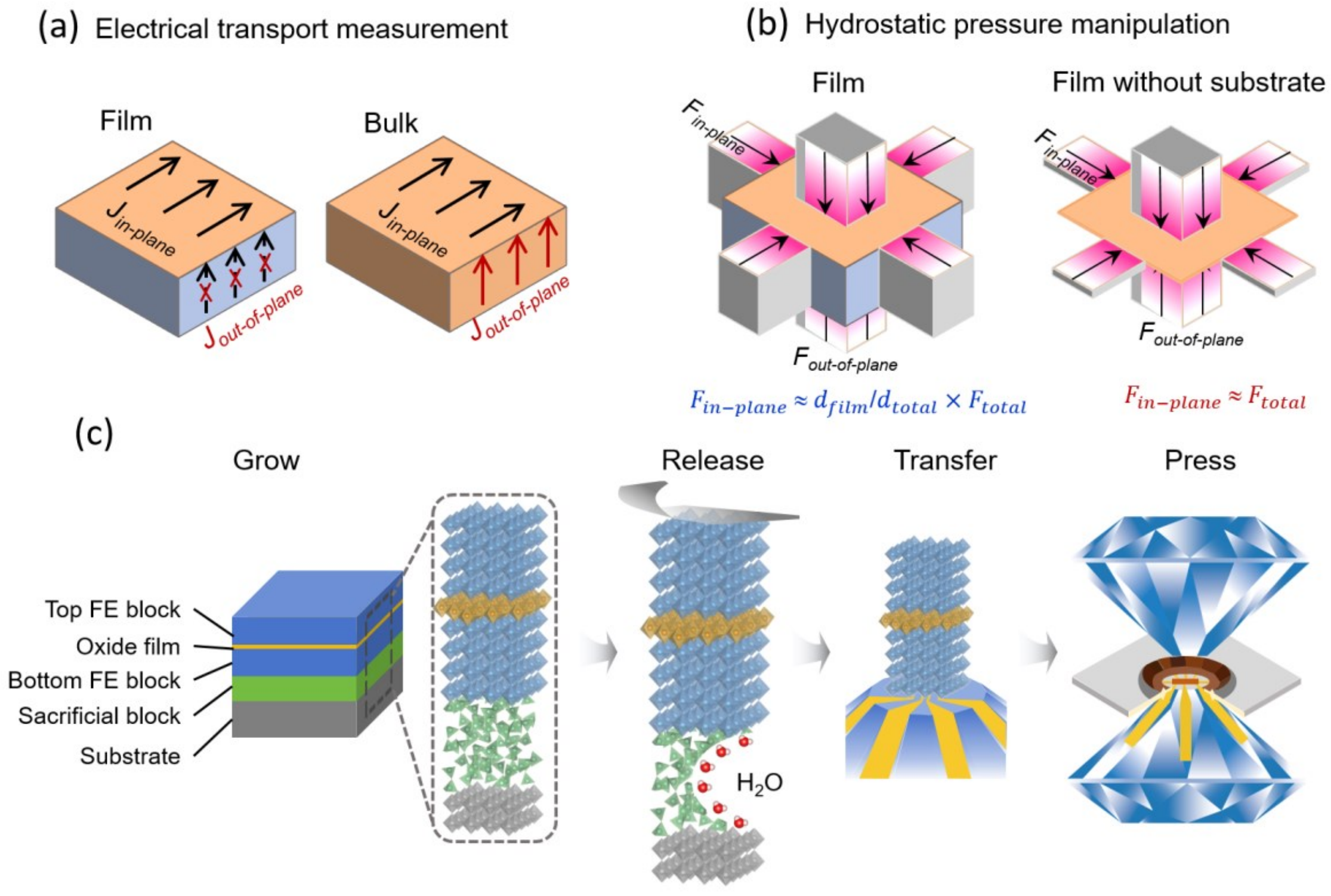

**Figure 1. (a)** Comparison of electrical transport measurements in thin-film and bulk samples. **(b)** Schematic illustration of the inefficiency of applying high pressure to thin films on a substrate. The in-plane force, $F_{\text{in-plane}}$, is determined by the ratio of the film thickness ($d_{\text{film}}$) to the total thickness ($d_{\text{total}}$). The total applied force is denoted by $F_{\text{total}}$. **(c)** Flowchart of the high-pressure strategy for freestanding thin films. The as-grown heterostructure consists of a sacrificial (SAO) layer, a bottom ferroelectric (FE) block, the oxide film, and a top FE block. The structure is released by dissolving the SAO layer in water and subsequently transferred into a DAC cell.

We showcased the merits of the encapsulated freestanding film in high-pressure studies by studying perovskite SIO for the following reasons, although the three-layer structure in principle works for almost all oxides. Firstly, SIO has a lattice parameter that is close to that of $BaTiO_3$ (BTO)[21]. The small lattice mismatch between SIO and BTO promotes coherent epitaxy in the layer-by-layer growth mode, which guarantees digital control of the thickness of each component[22]. Secondly, SIO is one of the well-known 5*d* materials, where the physical properties are determined by the delicate interplay between spin-orbit coupling and electronic correlation, and are expected to vary significantly under high pressure[23-26]. Thirdly, it is challenging to synthesize metastable perovskite SIO in bulk form[21], but it is relatively facile to stabilize this phase in thin films[27]. SIO is therefore a good example to showcase the importance of thin-film growth in materials science.

Explicitly, the three-layer structure is composed of an SIO film (30 u.c.) and two BTO blocks, as schematically shown in Fig. 1(c). The thickness of the BTO block was determined to be 5 u.c.[22], such that it is thick enough to reduce cracking but thin enough to prevent bending or twisting, as observed in pristine BTO membranes[13]. To release the three-layer structure from the substrate, a water-soluble $Sr_3Al_2O_6$ (SAO) block was first prepared on a single-crystal $SrTiO_3$ (STO) substrate prior to the growth of the three-layer structure. By etching the SAO block with water, an encapsulated freestanding SIO film that is robust against multiple transfer procedures was obtained. The freestanding film was then transferred into a DAC cell following the standard procedure for two-dimensional materials[7,19,20].

**Hydrostatic Pressure driven multiple phase transitions**

Figure 2(a) shows the XRD pattern of the encapsulated freestanding SIO film, where only Bragg reflections from the SIO layer are observed. The BTO layers are too thin to produce detectable peaks in the XRD pattern. The pronounced Kiessig fringes indicate high crystalline quality with sharp interfaces and flat surfaces. Notably, the typical sample size in DAC measurements is $50\times50$ $\mu m^2$. Therefore, obtaining a large-area, crack-free membrane is not strictly necessary for high-pressure studies, which significantly simplifies the transfer procedure. Nevertheless, as shown in the inset of Fig. 2(b), encapsulation is advantageous for obtaining sufficiently large and intact membranes compatible with DAC measurements.

We first confirmed that the freestanding film on the anvil exhibits a semimetallic state similar to that of films transferred onto a Si substrate under ambient conditions[22]. We then performed two independent pressure-dependent transport measurements on different pieces of the encapsulated freestanding SIO film. In the first DAC cell, with a maximum pressure of 6.0 GPa (Fig. 2(c)), the resistivity initially increases with pressure. Around 2.4 GPa, the resistivity increases by more than an order of magnitude, indicating a semimetal-to-insulator transition. With further increasing pressure, the resistivity decreases and is reduced by approximately four orders of magnitude at 6 GPa, indicating an insulator-to-metal transition. We refer to this state as a reentrant metallic phase to distinguish it from the semimetallic state at ambient pressure. Both transitions are reproduced in a second DAC cell within a similar pressure range (Fig. 2(d)). At higher pressures up to 16.5 GPa, the reentrant metallic phase exhibits a nontrivial evolution. We noted that the resistivity decreases with pressure up to ~9 GPa, above which it gradually increases.

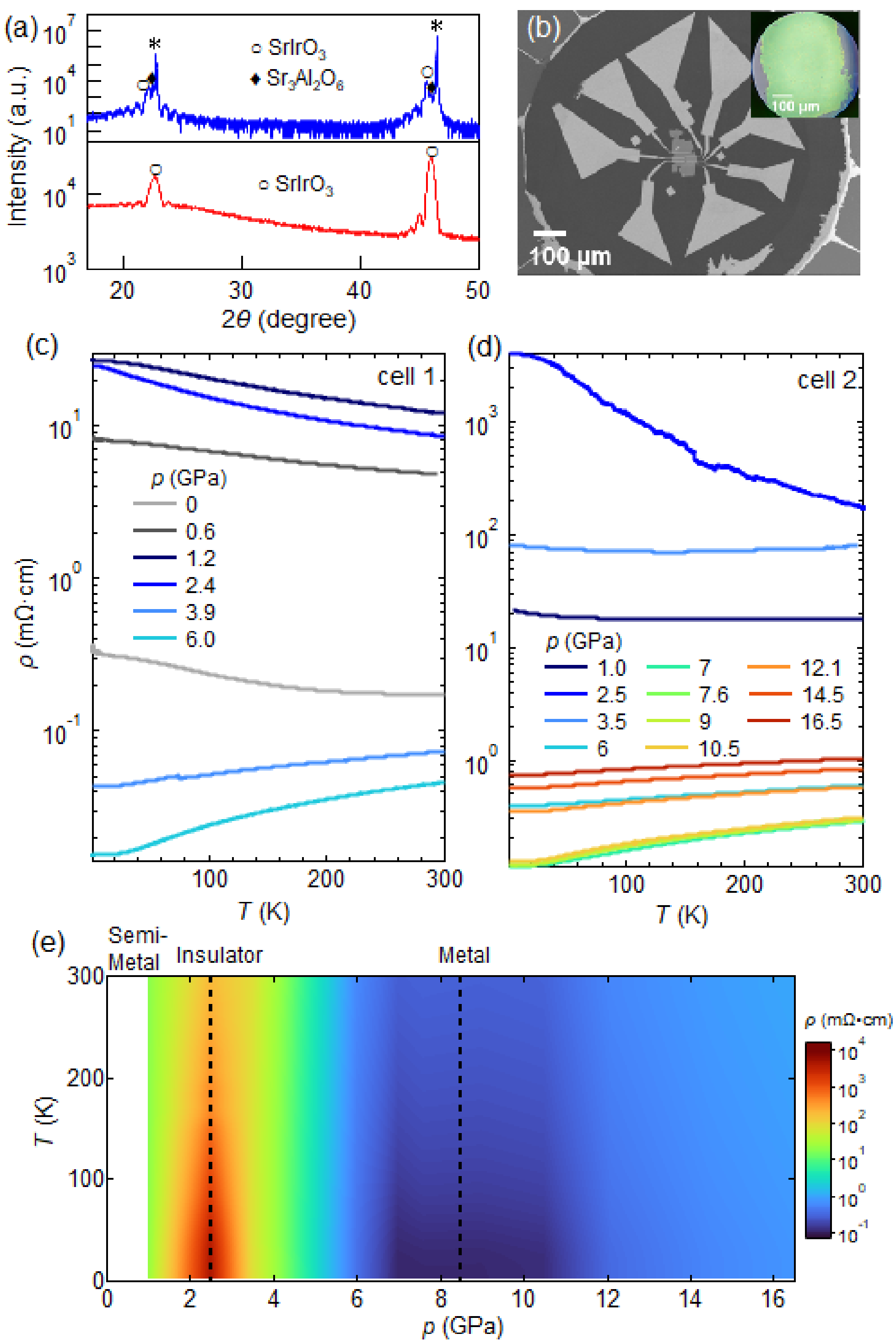

**Figure 2. (a)** XRD patterns of an epitaxial BTO/SIO/BTO film (top) and a freestanding BTO/SIO/BTO film (bottom). **(b)** Optical image of a DAC with patterned electrodes. Inset:

optical image of an encapsulated freestanding SIO film transferred onto a Si substrate. **(c&d)** Pressure-dependent resistivity of encapsulated freestanding SIO films. **(e)** Pressure-temperature phase diagram of the bulk-like SIO film derived from (d).

The pressure-dependent phase diagram is summarized in Fig. 2(e), where we determine two critical pressures around 2.5 and 9 GPa, at which the insulating and metallic behaviors are most pronounced, respectively. It is worth noting that pressure-driven multiple transitions have not been observed in SIO single crystals because the maximum applied pressure is only about 0.91 GPa[28]. Similarly, a persistent Dirac semimetal state has been observed in perovskite $CaIrO_3$ single crystals[29], where the maximum pressure reaches 3.2 GPa. While $CaIrO_3$ is a sister compound of perovskite SIO, its Fermi level lies closer to the Dirac point, and therefore the critical pressure for the semimetal-to-insulator transition may be slightly different from that of SIO[30]. We also note that the semimetal state remains robust in SIO thin films even when the pressure is increased to 2.65 GPa[31], around which a pronounced insulating state is observed in freestanding films (Fig. 2(e)). This comparison explicitly demonstrates that it is necessary to detach thin films from the substrate in order to maximize the effect of hydrostatic pressure.

On the other hand, the plenty of works on SIO epitaxial thin films[26] also enable an instructive comparison between the effects of hydrostatic pressure and epitaxial strain. It has been demonstrated that the semimetal-to-insulator transition can also be realized by applying a compressive strain of about -2.5% to SIO[32,33]. Phenomenologically, one may therefore assume that a compressive strain of -2.5% plays a role similar to that of a pressure of 2.5 GPa in triggering the semimetal-to-insulator transition. Following this direction, to induce the insulator-to-metal transition at 9 GPa in an epitaxial film, a compressive strain as large as -9% would be required. This value, however, is far beyond the typical epitaxial strain (~-3%) achievable by thin-film growth[18,34]. In other words, hydrostatic pressure of around 10 GPa represents an unmatched stimulus for uncovering emergent states in thin films.

**Capabilities on pressurizing monolayer-thick films and different compounds**

The most fascinating nature of thin-film growth is probably the flexible control of film thickness and chemical composition, which can be easily exploited in the high-pressure manipulation route. As a showcase, we have prepared a monolayer-thick SIO film that is encapsulated by two BTO blocks of 4 u.c., with the overall stacking order guaranteed by the RHEED intensity oscillations (Fig. 3(a)). The pressure-dependent transport properties of the encapsulated monolayer-thick SIO film were then systematically investigated by adopting the developed high-pressure strategy.

As shown in Fig. 3(b), the monolayer-thick SIO film displays a pronounced insulating nature at ambient pressure, similar to the reported insulating state in the ultrathin limit of SIO films and SIO/$SrTiO_3$ superlattices[35-38]. While the details of the lattice geometry are different, it is well established that both ultrathin films and superlattice structures lead to reduced dimensionality, which effectively enhances the electronic correlation in iridates and drives SIO from the semimetal phase to an antiferromagnetic insulating phase[35,37,39]. After applying hydrostatic pressure, we observe that the insulating state is weakened in a nonmonotonic manner, but ultimately remains stable even at 5.5 GPa. Note that the total thickness of the encapsulated freestanding film is only around 3.5 nm, and it easily breaks into pieces under pressure. As a result, we were not able to collect a convincing transport dataset beyond 5.5 GPa. Nonetheless, the pressure-robust insulating state in the two-dimensional limit is still strikingly different from the collapsed insulating state at ~2.5 GPa in the bulk-like film.

In addition to the spatially well-defined two-dimensional limit in the monolayer-thick film, $Sr_2IrO_4$ and $Ba_2IrO_4$ are also believed to be the two-dimensional counterparts of SIO, considering the separated $IrO_6$ single layers by the rock-salt SrO block[26]. Both $Sr_2IrO_4$ and $Ba_2IrO_4$ display a pronounced insulating state at ambient pressure due to the intriguing interplay between electronic correlation and spin-orbit coupling[40,41]. While $Ba_2IrO_4$ undergoes an insulator-to-metal transition at 13.8 GPa[42] and $Sr_2IrO_4$ preserves the insulating state under high pressure[43-45], it is clear that both bulk counterparts have the similar robust insulating state as that in the monolayer-thick SIO film under a moderate pressure value.

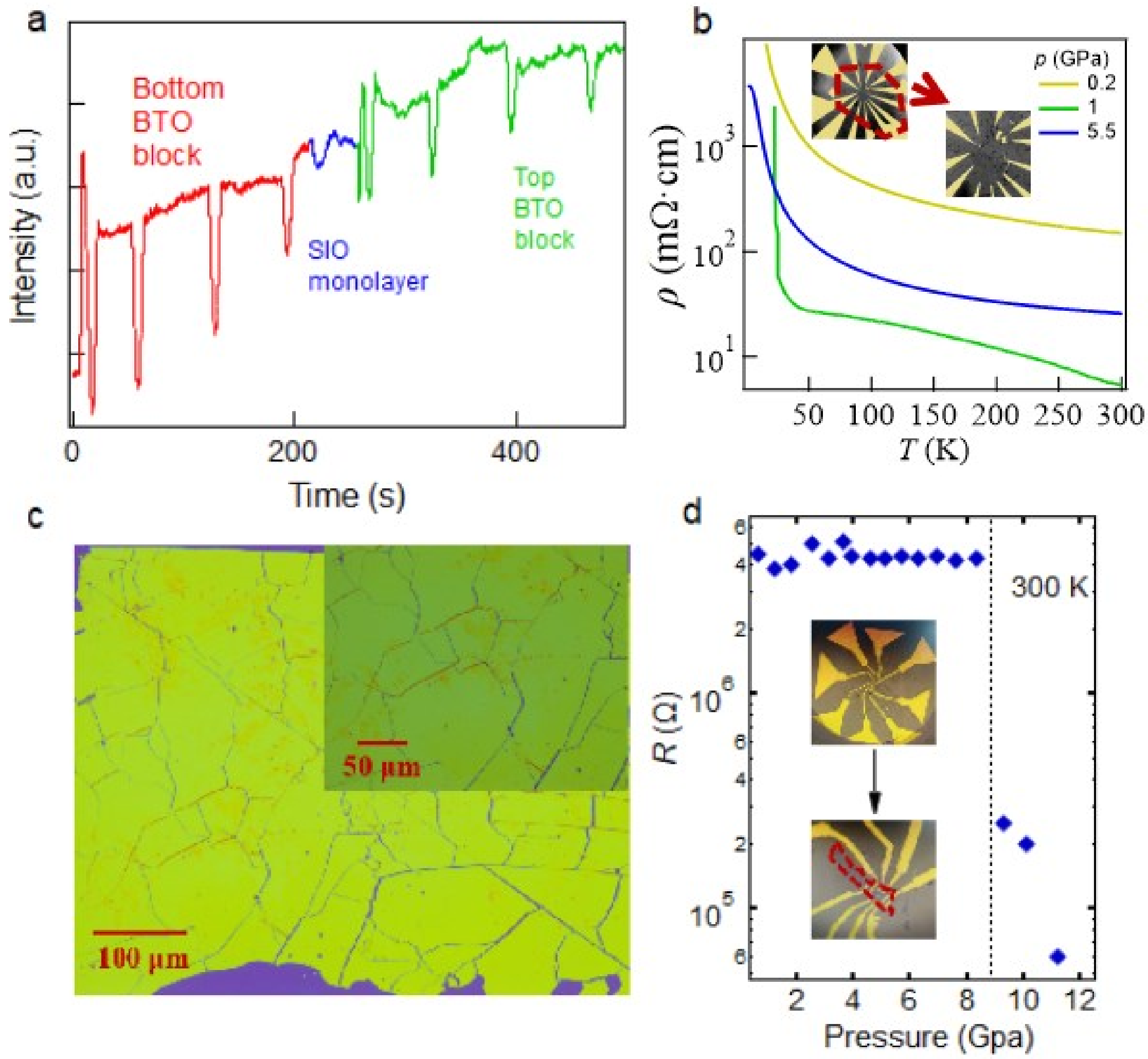

**Figure 3. (a)** RHEED intensity oscillations of the three-layer 4BTO/1SIO/4BTO structure grown on the soluble SAO layer. **(b)** Pressure-dependent electrical resistance of a monolayer SIO freestanding film. Inset: optical image of the patterned device in a DAC cell. A standard four-point method was used at 0.2 GPa, while a three-point method was employed at 1 and 5.5 GPa. **(c)** Optical image of a representative encapsulated freestanding STO film. **(d)** Pressure-dependent electrical resistance of the encapsulated STO freestanding film.

After demonstrating the high-pressure strategy with SIO films, we next show its applicability to different compounds. Here we focus on STO, which is another prototypical compound among perovskite oxides. Similarly, we first prepared an encapsulated STO freestanding film[22]. As shown in Fig. 3(c), the typical size of the capsule pieces is also sufficiently large to measure electrical transport in a DAC cell. This, again, showcases the advantage of the high-pressure strategy, *i.e.*, effectively avoiding the notorious difficulty in obtaining a millimeter-scale piece from a large class of freestanding films.

As summarized in Fig. 3(d), the encapsulated freestanding STO film displays good insulating behavior around ambient pressure, consistent with the band-insulator nature of titanates. In contrast to the significant modulation of electrical resistivity in SIO under high pressure, the electrical resistance of STO changes only slightly with hydrostatic pressure below 9 GPa (Fig. 3(d)). The resistance sharply decreases by one order of magnitude and displays a discontinuity at around 9 GPa, above which it decreases more rapidly with further increasing pressure. This observation is consistent with studies on STO single crystals, which reveal a cubic-to-tetragonal phase transition at pressures slightly below 10 GPa and a sudden drop in room-temperature resistance at the transition[46]. The consistency between high-pressure studies on freestanding thin films and single crystals reinforces the great potential of the high-pressure strategy for uncovering pressure-driven phenomena in compounds that can only be stabilized in thin-film form.

**Conclusion and outlook**

In this work, we present a universal strategy to probe the electrical transport of thin films under high pressure by integrating the substrate-free advantages of freestanding films with advanced high-pressure techniques on Nano-devices. We show that the hydrostatic pressure can reach as high as 16.5 GPa and is fully compatible with the standard four-point transport measurement method. We further demonstrate that the electrical transport of most thin films, even those as thin as a single monolayer, can be detected under high pressure if properly encapsulated by ferroelectric blocks. Using this high-pressure strategy, we discover a semimetal-to-insulator transition and an insulator-to-metal transition in bulk-like SIO thin films, while a robust insulating state is observed in the monolayer-thick SIO thin film. This observation indicates that high-pressure stimuli dictate emergent electronic states in SIO due to modulated electronic correlation.

Considering the readily obtainable high-pressure stimulus with the in-house apparatus, this strategy opens the door to investigating pressure-driven emergent electronic states in thin films, which remain largely unexplored due to the aforementioned experimental challenges. For example, it has been reported that $T_C$ of infinite-layer nickelate thin films increases almost linearly with pressure and does not saturate even when the pressure reaches as high as 12 GPa[11]. This observation raises an interesting question of how much $T_C$ would increase under further compression[47-50]. We note that robust superconductivity has recently been realized in freestanding nickelate thin films[51,52]. Therefore, an interesting direction is to investigate the pressure dependence of $T_C$ in freestanding superconducting films using the developed high-pressure strategy, with which pressures as high as 20 GPa may be achievable. Another potential direction is to discover and elucidate emergent phenomena in oxide films or heterostructures through high-pressure manipulation, including, but not limited to, two-dimensional superconductivity at oxide interfaces[53,54] and topologically protected electrical transport in heavy-metal oxide films, such as $SrRuO_3$[55].

*Comment*: We note that a recent study investigated thick freestanding SIO films under high pressure using X-ray absorption spectroscopy at room temperature. Although this work did not address the technical challenges associated with high-pressure electrical transport measurements on freestanding films, it nonetheless highlights the importance of applying high-pressure stimuli to freestanding oxide systems[56].

**Acknowledgement**

We appreciate helpful discussions with Lingfei Wang and Hong Zhu. This work was supported by National Key Basic Research Program of China (2024YFA1611100, 2022YFA1602603), National Natural Science Foundation of China (12104460, 12104459 and 12474125), the international partnership program of the Chinese Academy of Sciences (145GJHZ2022044MI), the HFIPS Director's Fund with Grant No. 2024YZGH04 and BJPY2024B07, the Anhui Provincial Major S&T Project (s202305a12020005) and the Basic Research Program of the Chinese Academy of Sciences Based on Major Scientific Infrastructures (JZHKYPT-2021-08).

**Competing interests:** The authors declare no competing financial interests.

**Data and materials availability:** All data in the main text or the supplementary materials are available upon reasonable request.